

\magnification=\magstep1
\footline={\hfill\ -- \folio\ -- \hfill}

\font\tenrm=cmr10     scaled 830
\font\tenit=cmti10    scaled 830
\font\elevenbf=cmbx10
\font\elevenrm=cmr10
\font\elevenit=cmti10

\font\ninerm=cmr9     scaled 830

\line{\hfil }
\hsize=6.0truein
\vsize=9.0truein


\rightline{NBI-HE-92-71}
\baselineskip=12truept
\rightline{KOBE-92-13}
\baselineskip=12truept
\rightline{October 1992}
\medskip

\centerline{\elevenbf PECULIARITY OF STRING THEORY ON ORBIFOLDS}
\medskip
\centerline{\elevenbf IN THE PRESENCE OF }
\medskip
\centerline{\elevenbf AN ANTISYMMETRIC BACKGROUND FIELD\footnote
            {$^\dagger$}{\ninerm\baselineskip=11pt
            To appear in the proceedings of the International Workshop
            on \lq\lq String Theory, Quantum Gravity and the
            Unification of the Fundamental Interactions", Rome,
            Italy on September 21-26, 1992.}}

\vglue 1.0truecm
\centerline{\elevenrm MAKOTO SAKAMOTO\footnote{$^\ddagger$}
{\ninerm\baselineskip=11pt On leave from Department of Physics, Kobe
University, Nada, Kobe 657, Japan}}

\baselineskip=13truept
\centerline{\tenit The Niels Bohr Institute, University of Copenhagen}
\baselineskip=12truept
\centerline{\tenit Blegdamsvej 17, DK-2100 Copenhagen X, Denmark}
\vglue 0.8truecm
\centerline{\elevenrm ABSTRACT}
\vglue 0.3truecm
{\rightskip=3truepc
 \leftskip=3truepc
 \tenrm\baselineskip=12truept
 \noindent
We study string theory on orbifolds in the presence of an antisymmetric
constant background field and discuss some of new aspects of the theory.
It is shown that the term with the antisymmetric field has a topological
nature like a Chern-Simons term or a Wess-Zumino term.
Due to this property,
the theory exhibits various anomalous behavior: Zero mode variables
obey nontrivial quantization conditions. Coordinate transformations
which define orbifolds are modified at the quantum level.
Wavefunctions of twisted strings acquire phase factors
when they move around
non-contractible loops on orbifolds. Zero mode eigenvalues are shifted
from naively expected values, in favor of modular invariance.
\vglue 0.8truecm }

\line{\elevenbf 1. Introduction \hfil}
\vglue 0.4truecm
\baselineskip=14truept

Geometrical notion of string theory is far from obvious.
For instance in toroidal compactification we can obtain enhanced
gauge symmetries by tuning the moduli of the torus${}^{1,2}$ and there
exists a striking isomorphism between the large radius and small
radius tori${}^3$. Another surprising example is that strings on some
orbifolds${}^4$ can equivalently be described as strings
on tori${}^{5-8}$
although orbifolds are geometrically quite different from tori.
An antisymmetric background field has first been introduced by
Narain, Sarmadi and Witten${}^9$ to explain Narain torus
compactification${}^{10}$ in the conventional canonical approach.
Orbifold models based on Narain torus compactification have
been discussed by many authors and various \lq\lq realistic"
models have been proposed${}^{11-14}$. Nevertheless, one interesting
topological nature of the antisymmetric field has not been
discussed before. Our aim of this paper is to discuss some of new
aspects of string theory on orbifolds in the presence of the
antisymmetric background field. As we will explain below, orbifold
models considered before are {\elevenit trivial}
in a topological point of view.
Our concern of this paper is topologically
{\elevenit nontrivial} orbifolds.
Since fermionic degrees of freedom play no important role in our
discussion, we will restrict our considerations to bosonic strings.

The action from which we shall start is${}^9$
$$
S[X]=\int d\tau \int_{0}^{\pi} d\sigma {1\over2\pi}\{\eta^{\alpha \beta}
\partial _{\alpha} X^I \partial_{\beta}X^I+B^{IJ}\varepsilon^{\alpha
\beta}\partial_{\alpha}X^I \partial_{\beta}X^J\},
\eqno(1.1)
$$
\noindent
where $B^{IJ}$\ $(I,J=1,\cdots,D)$ is the antisymmetric constant
background field and $X^I(\tau,\sigma)$\ $(I,\cdots,D)$ describes a
string coordinate on a $D$-dimensional torus $T^D$ which is defined
by identifying a point $X^I$ to $X^I+\pi w^I$ for all $w^I \in
\Lambda$, where $\Lambda$ is a $D$-dimensional lattice. An orbifold
${}^4$
is obtained by dividing a torus by the action of a discrete symmetry
group $G$ of the torus. Any element $g$ of $G$ can be represented
(for symmetric orbifolds) by${}^{4,15}$
$$
g = ( U , v ) ,
\eqno(1.2)
$$
where $U$ denotes a rotation and $v$ is a translation. To simplify
the discussion we will set the shift vector $v=0$ throughout this
paper. The shift vector will play no essential role in our
discussion. In section 2 we clarify a topological nature of the
second term of the action (1.1): Consider a transformation,
$$
g :\quad X^I\  \longrightarrow\  U^{IJ}X^J .
\eqno(1.3)
$$
It is shown in the next section that a Euclidean action $S_E[X]$
in the path integral formalism${}^{16}$ is not invariant under the
transformation (1.3) unless the rotation matrix $U^{IJ}$ commutes
with $B^{IJ}$ but in general
$$
S_E[X] - S_E[UX] = i2\pi n \quad \hbox{\elevenrm with}\ \  n \in
\hbox{\elevenbf Z} .
\eqno(1.4)
$$
This means that the transformation (1.3) is not a symmetry of the
Euclidean action $S_E[X]$ unless $[U,B]=0$ but is a symmetry of the
theory at the quantum level. Thus, th second term of the action (1.1)
can be regarded as a topological term like a Chern-Simons term or a
Wess-Zumino term. Much attention of previous works
has been paid to the case
$$
[ U , B ] = 0 ,  \eqno(1.5)
$$
or simply $B^{IJ}=0$, i.e., to topologically \lq\lq trivial" orbifold
models. The generalization to the case
$$
[ U , B ] \ne 0 ,  \eqno(1.6)
$$
is not, however, trivial, as expected from the result (1.4). A naive
construction of orbifold models with Eq.(1.6) would give modular
non-invariant partition functions and also destroy the duality of
amplitudes${}^{15,17,18}$.

Due to the property (1.4), the orbifold models with Eq.(1.6) exhibit
various anomalous behavior. In section 3 we discuss the duality of
amplitudes and clarify cocycle properties of vertex operators. Then
we find that zero modes of strings should obey nontrivial quantization
conditions, which are the origin of the anomalous behavior of the
theory in an operator formalism point of view.

In section 4 we see that the (classical) transformation should be
modified at the quantum level and that a vertex operator $V(k_L,k_R;z)$
transforms as
$$
g:\quad V(k_L,k_R;z)\ \  \longrightarrow \ \ \rho\  V(U^T k_L,U^T k_R;z).
  \eqno(1.7)
$$
The phase factor $\rho$ can be regarded as a kind of a quantum effect
and plays an important role in extracting physical states. It is pointed
out that in an algebraic point of view the phase $\rho$ has a connection
with automorphisms of algebras.

In section 5 we find an Aharonov-Bohm${}^{19}$ like effect in our system.
Strings on orbifolds in the presence of the antisymmetric background
field are very similar to electrons in the presence of an infinitely
long solenoid: The space is not simply-connected. The antisymmetric
background field plays the same role as an external gauge
field. If an electron moves around the solenoid, a wavefunction of the
electron in general acquires a phase. The same thing happens to twisted
strings.
A wavefunction of a twisted string is not, in general, periodic with
respect to a torus shift but
$$
\Psi(x^I+\pi w^I) = e^{i\pi w\cdot v}\Psi(x^I) ,  \eqno(1.8)
$$
for $w^I \in \Lambda$ such that $w^I = U^{IJ}w^J$. The constant shift
vector $v^I$ depends on the commutator $[U,B]$. We also discuss a
physical implication of Eq.(1.8) and see how a naively expected spectrum
is modified.

Section 6 is devoted to conclusions.

\vglue 0.6cm

\line{\elevenbf 2. Topological Nature of a ${\bf B^{IJ}}$-Term \hfil}
\vglue 0.4cm

In this section, we shall clarify a topological nature of the second term
of the action (1.1) from a path integral point of view.

In the Euclidean path integral formalism, the one-loop vacuum amplitude
of the closed bosonic string theory on a torus is given by the functional
integral${}^{16}$,
$$
\int {[dg_{\alpha \beta}][dX^I]\over{\cal V}_{ol}}\  \exp \{ -S[X,g] \} ,
\eqno(2.1)
$$
where $g_{\alpha \beta}$ is a Euclidean metric of the two dimensional
world sheet with the topology of a torus and ${\cal V}_{ol}$ is a volume
of local gauge symmetry groups. The Euclidean action is given by
$$
S[X,g]=\int_0^1 d^2\sigma {1\over2\pi}\{\sqrt g g^{\alpha \beta}
\partial _{\alpha} X^I \partial_{\beta}X^I
-iB^{IJ}\varepsilon^{\alpha
\beta}\partial_{\alpha}X^I \partial_{\beta}X^J\}. \eqno(2.2)
$$
It should be emphasized that the imaginary number $i$ appears in the
second term of the Euclidean action (2.2) due to the antisymmetric
property  of $\varepsilon^{\alpha \beta}$. Since strings propagate on
a torus defined by identifying a point $X^I$ with $X^I+\pi w^I$ for
all $w^I \in \Lambda$, where $\Lambda$ is a $D$-dimensional lattice,
the string coordinate $X^I(\sigma^1,\sigma^2)$ obeys the following
boundary conditions:
$$
\eqalign{X^I(\sigma^1+1,\sigma^2) &= X^I(\sigma^1,\sigma^2) + \pi w^I ,\cr
         X^I(\sigma^1,\sigma^2+1) &= X^I(\sigma^1,\sigma^2) + \pi{w'}^I ,
         }
\eqno(2.3)
$$
for some $w^I,{w'}^I \in \Lambda$.

Let us consider a transformation,
$$
g :\quad X^I\  \longrightarrow \ U^{IJ}X^J \quad
\hbox{\elevenrm with}\ \
U^TU = \hbox{\elevenbf 1},
\eqno(2.4)
$$
where $U^{IJ}$ is an orthogonal matrix. The first term of the action
(2.2) is trivially invariant under the transformation (2.4) but the
second term is not invariant if $U^{IJ}$ does not commute with $B^{IJ}$.
Since $B^{IJ}$ is a constant antisymmetric field, the second term of the
action (2.2) can be written as a total divergence, i.e.,
$$
B^{IJ} \varepsilon^{\alpha \beta}\partial_\alpha X^I
\partial_\beta X^J = \partial_\alpha(B^{IJ}\varepsilon^{\alpha \beta}
X^I \partial_\beta X^J).
\eqno(2.5)
$$
It turns out that the difference between $S[X,g]$ and $S[UX,g]$ is
given by
$$
S[X,g] - S[UX,g] = i\pi {w'}^I (B-U^TBU)^{IJ}w^J .
\eqno(2.6)
$$
{}From this relation, we come to an important conclusion that the
transformation (2.4) is a {\elevenit quantum} symmetry of the
theory in a path integral point of view if
$$
(B-U^TBU)^{IJ}w^J \in 2\Lambda^* \quad \hbox{\elevenrm for\  all}\ \
w^I \in \Lambda,
\eqno(2.7)
$$
where $\Lambda^*$ is the dual lattice of $\Lambda$, although the
Euclidean action (2.2) itself is not invariant under the transformation
(2.4) if $[U,B]\ne 0$.
Then, the second term of the action (2.2)
can be regarded as a topological term like a Chern-Simons term or a
Wess-Zumino term.
In the following, we will show that
the condition (2.7) is nothing but a consistency condition for
constructing string theory on orbifolds.

An orbifold ${}^{4}$ is obtained by dividing a torus by the action
of a suitable discrete group $G$.
In the construction of an orbifold model, we start with a $D$-dimensional
toroidally compactified closed bosonic string theory which is
specified by a $(D+D)$-dimensional Lorentzian even self-dual lattice
$\Gamma^{D,D^{9,10}}$, on which the left- and right-moving momentum
$(p_L^I,p_R^I)$ $(I=1,\cdots,D)$ lies.
In general any group element $g$ of $G$ (for symmetric orbifolds)
can be represented by
$$
g = (U,v),
\eqno(2.8)
$$
where $U$ is a rotation and $v$ is a translation.
As mentioned in the introduction, we will set the shift vector
$v=0$ for simplicity.
To define an orbifold consistently, $G$ must be a discrete symmetry
group of the torus.
This means that any element $g$ of $G$ must be an automorphism
of the lattice $\Gamma^{D,D}$,i.e.,
$$
g:\quad (p^I_L,p^I_R)\  \longrightarrow\  (U^{IJ}p^J_L,U^{IJ}p^J_R)
\in \Gamma^{D,D}.
\eqno(2.9)
$$
for all $(p_L^I,p_R^I)\in\Gamma^{D,D}$.
The left- and the right- moving momenta, $p_L^I$ and $p_R^I$,
are related to the center of mass momentum $p^I$ and the winding
number $w^I$ of a string as follows${}^9$:
$$
\eqalign{
p^I_L &={1\over2}p^I+{1\over2}(1-B)^{IJ}w^J, \cr
p^I_R &={1\over2}p^I-{1\over2}(1+B)^{IJ}w^J. \cr
}
\eqno(2.10)
$$
The winding number $w^I$, by definition, lies on the lattice
$\Lambda$, i.e.,
$$
w^I \in \Lambda.
\eqno(2.11)
$$
Since a wavefunction $\Psi(x^I)$ of a string on a torus
must be periodic, i.e.,
$\Psi(x^I+\pi w^I)=\Psi(x^I)$ for all $w^I\in\Lambda$,
the allowed momentum is
$$
p^I \in 2\Lambda^*.
\eqno(2.12)
$$
It follows from Eqs. (2.9) and (2.10) that $g$ acts on
$w^I$ and $p^I$ as follows:
$$
\eqalign{
g:\quad w^I\  &\longrightarrow \ U^{IJ}w^J,  \cr
        p^I\  &\longrightarrow \ U^{IJ}p^J - [U,B]^{IJ}w^J. \cr
}
\eqno(2.13)
$$
For the action of $g$ on $w^I$ to be well-defined, it must be
an automorphism of $\Lambda$,i.e.,
$$
U^{IJ}w^J \in \Lambda \quad
\hbox{\elevenrm for\ all}\ \  w^I \in \Lambda.
\eqno(2.14)
$$
For the action of $g$ on $p^I$ to be well-defined, we further
require the following condition:
$$
[U,B]^{IJ}w^J \in 2\Lambda^*,
$$
or equivalently
$$
(B-U^TBU)^{IJ}w^J \in 2\Lambda^* \quad \hbox{\elevenrm for\  all}\ \
w^I \in \Lambda.
\eqno(2.15)
$$
This is just the condition (2.7), as announced before.

In the following sections, we shall construct the operator
formalism of string theory on orbifolds in the presence
of the antisymmetric constant background field and reveal
various anomalous behavior of the theory.
We will see that the necessity of nontrivial cocycle operators
in vertex operators is the origin of the anomalous behavior in
the operator formalism.

Before closing this section, it may be instructive to present
some examples of $B^{IJ}$ and $U^{IJ}$,
which satisfy the condition (2.15).
Let us consider the following $(D+D)$-dimensional Lorentzian
even self-dual lattice, which has been introduced by
Englert and Neveu${}^{20}$,
$$
\Gamma^{D,D} = \{\ (p^I_L,p^I_R)\ |\ p^I_L,p^I_R \in
\Lambda_W({\cal G}),\ p^I_L - p^I_R \in \Lambda_R({\cal G})\ \},
\eqno(2.16)
$$
where $\Lambda_R(\cal G)$ $(\Lambda_W(\cal G))$ is the root
(weight) lattice of a simply-laced Lie algebra $\cal G$
with rank $D$ and the squared length of the root vectors
is normalized to two.
In this normalization, the weight lattice $\Lambda_W(\cal G)$
is just the dual lattice of $\Lambda_R(\cal G)$.
It should be emphasized that the lattice (2.16) plays a crucial
role of gauge symmetries in closed string theory${}^{1,2}$.
The lattice (2.16) can be obtained through the relations (2.10)
by choosing $\Lambda$ and $B^{IJ}$ as follows:
$$
\Lambda = \Lambda_R(\cal G) ,
\eqno(2.17)
$$
and
$$
\alpha^I_i B^{IJ}  \alpha^J_j = \alpha^I_i \alpha^I_j
\quad \hbox{\elevenrm mod}\  2 ,
\eqno(2.18)
$$
where $\alpha_i$ is a simple root of $\cal G$ which is normalized
to $(\alpha_i)^2=2$.
If we choose the rotation matrix $U$ to be an automorphism of
the root lattice $\Lambda_R({\cal G})$, i.e.,
$$
U^{IJ}w^J \in \Lambda_R({\cal G}) \quad \hbox{\elevenrm for\ all}\ \
w^I \in \Lambda_R({\cal G}) ,
\eqno(2.19)
$$
then the transformation (2.9) is an automorphism of the lattice
(2.16) and
the matrix $U$ always satisfies the condition (2.15), i.e.,
$$
(B - U^TBU)^{IJ}w^J \in 2\Lambda_W({\cal G}) \quad
\hbox{\elevenrm for\  all}\ \  w^I \in \Lambda_R({\cal G}) .
\eqno(2.20)
$$

\vglue 0.6cm

\line{\elevenbf 3. Cocycle Properties of Vertex Operators \hfil}
\vglue 0.4cm

In this section we shall investigate cocycle properties of vertex
operators and show that zero modes of strings should obey nontrivial
quantization conditions.

Let us consider a vertex operator $V(k_L,k_R,z)$ which
describes the emission of a state with the momentum
$(k_L^I,k_R^I)\in\Gamma^{D,D}$.
The vertex operator will be of the form,
$$
V(k_L,k_R;z) = f_{k_L,k_R}(z)
:e^{ik_L\cdot X_L(z)+ik_R\cdot X_R(\bar z)}C_{k_L,k_R}:,
\eqno(3.1)
$$
where $X_L^I(z)$ $(X_R^I(\bar z))$ are the left- (right-)
moving string coordinate and $f_{k_L,k_R}(z)$ is a normalization
factor, which will in general depend on $k_L^I,k_R^I,z$ and
$\bar z$ in twisted sectors.
The $C_{k_L,k_R}$ is, if necessary, some extra (cocycle) operator.
The product of two vertex operators
$$
V(k_L,k_R;z)V(k'_L, k'_R;z'),
\eqno(3.2)
$$
is well-defined if $|z|>|z'|$.
The different ordering of the two vertex operators corresponds to
the different \lq\lq time"-ordering. To obtain scattering amplitudes,
we must sum over all possible \lq\lq time"-ordering for the emission
of states. We must then establish that each contribution is
independent of the order of the vertex operators to enlarge the
regions of integrations over $z$ variables$^{21}$. Thus the product (3.2),
with respect to $z$ and $z'$, has to be analytically continued to the
region $|z'|>|z|$ and to be identical to
$$
V(k'_L,k'_R;z')V(k_L, k_R;z),
\eqno(3.3)
$$
for $|z'|>|z|$.
In terms of zero modes, the above statement can be translated into the
following condition:
$$
V_0(k_L,k_R)V_0(k'_L,k'_R)= \eta\
V_0(k'_L,k'_R)V_0(k_L,k_R),
\eqno(3.4)
$$
where
$$
V_0(k_L,k_R)=e^{ik_L\cdot \hat x_L+ik_R\cdot \hat x_R}C_{k_L,k_R}.
\eqno(3.5)
$$
The $\hat x_L^I$ $(\hat x_R^I)$ denotes the left-(right-) moving
\lq\lq center of mass" coordinate.
The phase factor $\eta$ is required to compensate a phase appearing
in reversing the order of the oscillator modes of the vertex operators.
In the untwisted sector, the phase $\eta$ is given by${}^{1,2}$
$$
\eta = (-1)^{k_L\cdot k'_L - k_R\cdot k'_R}.
\eqno(3.6)
$$
In the $U$-twisted sector with $U^N={\bf 1}$, the phase $\eta$
is given by${}^{22}$
$$
\eta = \exp \{ i\pi k^I_L(1 + \sum^N_{\ell =1}{\ell \over N}
(U^{-\ell}-U^\ell))^{IJ}{k'}^J_L
- i\pi k^I_R(1 + \sum^N_{\ell =1}{\ell \over N}
(U^{-\ell}-U^\ell))^{IJ}{k'}^J_R\}.
\eqno(3.7)
$$

In the untwisted sector, Frenkel and Ka\v c${}^{1,2}$ have shown
that the cocycle operator can be constructed without introducing
any more degrees of freedom.
However, instead of introducing the cocycle operator
$C_{k_L,k_R}$ it is possible to achieve the relation (3.4).
In ref.23, it has been shown that multiplying vertex operators by
cocycle operators is equivalent to modifying commutation
relations for zero modes and that the following commutation
relations
\footnote{$^{\dagger}$}{\ninerm\baselineskip=11pt
The normalization of $\scriptstyle \hat x_L^I$ and
$\scriptstyle \hat x_R^I$ is different
from that in ref.23 by factor two.}
with $C_{k_L,k_R}=1$:
$$
\eqalign{
[\hat x^I_L,\hat x^J_L] &= i\pi B^{IJ}, \cr
[\hat x^I_R,\hat x^J_R] &= i\pi B^{IJ}, \cr
[\hat x^I_L,\hat x^J_R] &= i\pi (1-B)^{IJ}, \cr
}
\eqno(3.8)
$$
lead to the relation (3.4) with the correct phase (3.6).
A geometrical meaning of the above commutation relations has
also been discussed in ref.23.

In twisted sectors, the relation (3.4) with the phase (3.7)
may require new degrees of freedom, which are called fixed
points (fixed lines, fixed surfaces,$\cdots$)
by physicists.
The explicit realization has been obtained in ref.24 in the
case of $[U,B]=0$ and $det(1-U)\ne 0$.
In ref.24, the quantization of zero modes of twisted strings
has been clarified from a geometrical point of view
and the nontrivial commutation relations of the zero modes with
$C_{k_L,k_R}=1$ have been shown to naturally satisfy the relation
(3.4) with the correct phase (3.7).
For $[U,B]\ne 0$ and/or $det(1-U)=0$ we need to modify the
results given in ref.24.
It turns out that the following commutation relations with
$C_{k_L,k_R}=1$ satisfy the relation (3.4) with the phase (3.7):
$$
\eqalign{
[\hat x^I_L,\hat x^J_L] &= i\pi (B-\sum^N_{\ell = 1}
{\ell\over N}(U^{-\ell}-U^\ell))^{IJ}, \cr
[\hat x^I_R,\hat x^J_R] &= i\pi (B+\sum^N_{\ell = 1}
{\ell\over N}(U^{-\ell}-U^\ell))^{IJ}, \cr
[\hat x^I_L,\hat x^J_R] &= i\pi (1-B)^{IJ}, \cr
}
\eqno(3.9)
$$
These commutation relations reduce to those given in ref. 24 for
$[U,B]~=~0$ and $det(1-U)\ne 0$ and also reduce to the relations
(3.8) for $U=\hbox{\elevenbf 1}$.
We will see in the following sections that the above nontrivial
commutation relations are the origin of anomalous
features of the theory.

\vglue 0.6cm

\line{\elevenbf 4. Anomalous Transformations \hfil}
\vglue 0.4cm

In this section, we shall show that the action of $g$ on the string
coordinate becomes anomalous at the quantum level if $[U,B]\ne 0$.

The relevant operators in string theory are the momentum operators,
$P_L^I(z)$ \par
\noindent
$\equiv i\partial_zX_L^I(z)$ and
$P_R^I(\bar z)\equiv i\partial_{\bar z}X_R^I({\bar z})$,
and the vertex operator $V(k_L,k_R;z)$.
Other operators can be obtained from the operator products of these
operators.
For example the energy-momentum tensors of the left- and the right-
movers are given by
$$
\eqalign{
T(z)&=\lim_{w\to z}{1\over 2}\Bigl(P^I_L(w)P^I_L(z)-
{D \over (w-z)^2}\Bigr),
\cr
\bar T(\bar z)&=\lim_{\bar w\to \bar z}
{1 \over 2}\Bigl(P^I_R(\bar w)P^I_R(\bar z)-
{D \over (\bar w-\bar z)^2}\Bigr).
\cr
}
\eqno (4.1)
$$
Under the action of $g$, the momentum operators transform as
$$
g:\quad (P^I_L(z),P^I_R(\bar z))\  \longrightarrow \
(U^{IJ}P^I_L(z),U^{IJ}P^I_R(\bar z)),
\eqno(4.2)
$$
which leaves the energy-momentum tensors invariant,
as it should be.
One might expect that under the action
of $g$ the left- and the right- moving string coordinates,
$X_L^I(z)$ and $X_R^I({\bar z})$, transform in a similar
way to $P_L^I(z)$ and $P_R^I({\bar z})$.
However, it is not the case if $U^{IJ}$ does not commute with
$B^{IJ}$.
Suppose that under the action of $g$, ${\hat x}_L^I$ and
${\hat x}_R^I$ would transform as
$$
\eqalign{
g:\quad \hat x^I_L\  &\longrightarrow \ U^{IJ}\hat x^J_L, \cr
        \hat x^I_R\  &\longrightarrow \ U^{IJ}\hat x^J_R. \cr
}
\eqno(4.3)
$$
It is easy to see that the transformations (4.3) is
inconsistent with the commutation relations (3.8) or (3.9)
unless $U^{IJ}$ commutes with $B^{IJ}$.
In ref.15, it has been shown that in the untwisted sector the correct
action of $g$ on ${\hat x}_L^I$ and ${\hat x}_R^I$ is given by
\footnote{$^\dagger$}{\ninerm\baselineskip=11pt
Set $\scriptstyle U_L=U_R\equiv U$
and assume the form (6-13) in the first
reference of ref.15.}
$$
\eqalign{
g:\quad \hat x^I_L\  &\longrightarrow \ U^{IJ}\hat x^J_L
        + \pi U^{IJ}({1\over 2}(B-U^TBU)-C_U)^{JK}\hat w^K, \cr
        \hat x^I_R \ &\longrightarrow \ U^{IJ}\hat x^J_R
        - \pi U^{IJ}({1\over 2}(B-U^TBU)-C_U)^{JK}\hat w^K, \cr
}
\eqno(4.4)
$$
where ${\hat w}^k$ is the winding number operator,
which satisfies the following commutation relations${}^{24}$:
$$
\eqalign{
[\hat x^I_L,\hat w^J] &= i\delta^{IJ}, \cr
[\hat x^I_R,\hat w^J] &= -i\delta^{IJ}.\cr
}
\eqno(4.5)
$$
The $C_U^{IJ}$ is a symmetric matrix defined through the relation,
$$
w^IC_U^{IJ}w'^J = {1\over 2}w^I(B-U^TBU)^{IJ}w'^J \quad
\hbox{\elevenrm mod}\ 2,
\eqno(4.6)
$$
for all $w^I,{w'}^I\in \Lambda$.
The existence of a symmetric matrix $C_U$ is guaranteed by
the fact that
$$
{1\over 2}w^I(B-U^TBU)^{IJ}w'^J \in \hbox{\elevenbf Z} \quad
\hbox{\elevenrm for\ all}\ \ w^I,w'^I \in \Lambda.
\eqno(4.7)
$$
In ref.15, the transformation (4.4) has been derived from the
requirement that the action of $g$ on vertex operators has
to preserve the duality property of amplitudes.
It is important to verify that the transformation (4.4)
is consistent with the commutation relations (3.8).
We can further show that the transformation (4.4)
is still correct in the $U$-twisted sector and
is consistent with the commutation relations (3.9).

It seems that the transformation (4.4) loses its geometrical
meaning unless $U^{IJ}$ commutes with $B^{IJ}$.
However, the transformation (4.4) is geometrically still
well-defined because the anomalous terms in Eq.(4.4) can
be regarded as a torus shift, i.e.,
$$
g:\quad (\hat x^I_L,\hat x^I_R)\ \longrightarrow \
        (U^{IJ}\hat x ^J_L,U^{IJ}\hat x^J_R)
        + \hbox{\elevenrm torus\ shift}.
\eqno(4.8)
$$
To see this, let us consider the action of $g$ on
$k_L\cdot {\hat x}_L+k_R\cdot {\hat x}_R$ for
$(k_L^I,k_R^I)\in \Gamma^{D,D}$.
$$
g:\quad k_L\cdot \hat x_L + k_R\cdot \hat x_R\  \longrightarrow \
        k_L\cdot U\hat x_L + k_R\cdot U\hat x_R
       +\pi (k_L-k_R)\cdot U({1\over 2}(B-U^TBU)-C_U)\hat w.
\eqno(4.9)
$$
The last term of the right hand side may be regarded as a
trivial shift because
$$
(k_L-k_R)\cdot U({1\over 2}(B-U^TBU)-C_U)\hat w = 0
\quad \hbox{\elevenrm mod}\ 2,
\eqno(4.10)
$$
for all $(k_L^I,k_R^I)\in \Gamma^{D,D}$ and
${\hat w}^I\in \Lambda$.
(Recall that $k_L^I-k_R^I\in \Lambda$.)

What role do the anomalous terms in Eq.(4.4) play?
To answer this question, let us consider the action
of $g$ on a vertex operator.
The result is
$$
g:\quad V(k_L,k_R;z)\ \  \longrightarrow \ \
e^{-i{\pi \over 2}(k_L-k_R)\cdot UC_UU^T(k_L-k_R)}
\ V(U^T k_L,U^T k_R;z).
\eqno(4.11)
$$
It should be noted that to derive the relation (4.11)
we cannot use the relation (4.10) directly in the exponent
because ${\hat w}^I$ is not a $c$-number
but a $q$-number.
We can use the relation (4.10) only after separating
the terms depending on ${\hat w}^I$ from
$V(U^Tk_L,U^Tk_R;z)$ by use of the Hausdorff formula.
The phase factor appearing in the transformation (4.11)
plays an important role in extracting physical states,
which must be $g$-invariant.
The following example may be helpful to understand a
role of the phase in Eq.(4.11):
Let us consider the (left-moving) momentum operator
$H^I(z)\equiv P_L^I(z)$ $(I=1,2)$ and the vertex operators
$E^\alpha (z)\equiv V(\alpha ,0;z)$,
where $\alpha$'s are root vectors of $SU(3)$ with $\alpha ^2=2$.
They will form level one $SU(3)$ Ka\v c-Moody algebra${}^{1,2}$.
Let $\alpha_i$ $(i=1,2)$ be a simple root of $SU(3)$.
Consider the following transformation:
$$
g:\quad \alpha_1\  \leftrightarrow\  \alpha_2.
\eqno(4.12)
$$
This is clearly an automorphism of the root
lattice of $SU(3)$.
Then one might expect that under the transformation (4.12)
the generators of $SU(3)$ Ka\v c-Moody algebra would transform as
follows:
$$
\eqalign{
\alpha_1\cdot H(z)\  &\longleftrightarrow \ \alpha_2\cdot H(z), \cr
E^{\pm\alpha_1}(z)\  &\longleftrightarrow \ E^{\pm\alpha_2}(z), \cr
}
\eqno(4.13)
$$
and $E^{\pm (\alpha_1+\alpha_2)}(z)$ would be left invariant.
If it were ture, we would have a strange result:
The invariant generators, $(\alpha_1+\alpha_2)\cdot H(z)$,
$E^{\pm\alpha_1}(z)+E^{\pm\alpha_2}(z)$ and $E^{\pm(\alpha_1+\alpha
_2)}(z)$  could form a subalgebra of $SU(3)$.
This is not, however, acceptable because five generators must
form an algebra with rank more than 3 while
rank of $SU(3)$ is 2.
The key to resolve this inconsistency is the phase in Eq.(4.11):
We may take an Englert-Neveu lattice (2.16) with
$\Lambda_R(SU(3))$ in order for $E^\alpha(z)$ to be well-
defined${}^{1,2}$.
This means that we take a nontrivial antisymmetric field $B^{IJ}$
through the relation (2.18).
It turns out that $B^{IJ}$ does not commute with the transformation
(4.12).
Then, our formalism tells us that the transformations (4.13)
are still true but $E^{\pm(\alpha_1+\alpha_2)}(z)$ must
transform as
$$
g:\quad E^{\pm (\alpha_1+\alpha_2)}(z)
\ \longrightarrow \ -E^{\pm (\alpha_1+\alpha_2)}(z).
\eqno(4.14)
$$
Therefore, $E^{\pm(\alpha_1+\alpha_2)}(z)$ are not invariant
under the action of $g$ although $\alpha_1+\alpha_2$ is.
In an algebraic point of view the phase in Eq.(4.11) has
a connection with automorphisms of algebras rather than
automorphisms of lattices.

\vglue 0.6cm

\line{\elevenbf 5. Aharonov-Bohm like Effect \hfil}
\vglue 0.4cm

Strings on orbifolds in the presence of the antisymmetric
background field may be in a similar situation to
electrons in the presence of an infinitely long solenoid.
Both underlying spaces are not simply-connected and
possess non contractible loops.
The antisymmetric background field $B^{IJ}$ may play a similar
role to an external gauge field $A_\mu$.
The gauge field $A_\mu$ is not gauge invariant and
$B^{IJ}$ is not invariant under the transformation
$B^{IJ}$ $\rightarrow$ $(U^TBU)^{IJ}$ if $U^{IJ}$ does
not commute with $B^{IJ}$.
The relation (2.6) means that $B^{IJ}$ cannot
be defined as a single-valued \lq\lq function" for twisted strings
with $[U,B]\ne 0$.
This fact suggests that if a twisted string moves around
a non-contractible loop on the orbifold the wavefunction
may acquire a nontrivial phase.
In fact we can show that
$$
\Psi(x^I+\pi w^I) = e^{-i{\pi\over 2} w^I C_U^{IJ}w^J}\Psi(x^I) ,
\eqno(5.1)
$$
where
$$
w^I \in \Lambda_U = \{w^I \in \Lambda\  |\  w^I = U^{IJ}w^J\}.
\eqno(5.2)
$$
What is a physical implication of the relation (5.1) ?
To see this, we first note that the left hand side of Eq.(5.1)
can be expressed as
$$
\Psi(x^I+\pi w^I) = e^{-i\pi w\cdot \hat p_{{}_{/\negthinspace/}}}
\Psi(x^I) ,
\eqno(5.3)
$$
where $\hat p^I_{{}_{/\negthinspace /}}$ is the canonical
conjugate momentum
restricted to the $U$-invariant subspace, i.e.,
$\hat p^I_{{}_{/\negthinspace /}}=
U^{IJ}\hat p^J_{{}_{/\negthinspace /}}$.
Introduce a vector $v^I$ with $v^I=U^{IJ}v^J$ through the relation
$^{15}$
$$
w^Iv^I = {1\over 2}w^IC_U^{IJ}w^J \quad \hbox{\elevenrm mod}\ 2,
\eqno(5.4)
$$
for all $w^I \in \Lambda_U$. Comparing Eq.(5.1)
with (5.3), we conclude that
$$
\hat p^I_{{}_{/\negthinspace /}} \in v^I+{\Lambda_U}^*,
\eqno(5.5)
$$
where ${\Lambda_U}^*$ is the dual lattice of $\Lambda_U$.
Thus, allowed eigenvalues of
$\hat p^I_{{}_{/\negthinspace /}}$ are
different from naively expected values$^6$ by $v^I$.
This result exactly agrees with the one expected from
the argument of modular invariance in ref.15.
Further we can show that Eq.(5.5) is consistent with
the single-valuedness of $g$-invariant operators with
respect to $z$.

Before deriving Eq.(5.1) or (5.5), it may be instructive to recall
quantum mechanics on a circle with a radius $L$,
where a point $x$ is identified with $x+2\pi nL$ for all
$n\in {\bf Z}$.
It is important to understand that the coordinate operator
${\hat x}$ itself is not well-defined on the circle while
the canonical momentum operator $\hat p$ is well-defined.
However, the following operator
$$
e^{ik\hat x} \quad \hbox{\elevenrm with}\ \ k={m\over L}\ \ (m\in
\hbox{\elevenbf Z}),
\eqno(5.6)
$$
is well-defined on the circle because it is consistent with the
torus identification.
Physically, the operator (5.6) plays a role of a momentum shift
by $k={m\over L}$.
It turns out that an operator $e^{i2\pi L{\hat p}}$ commutes
with all operators, i.e.,
${\hat p}$ and $e^{ik{\hat x}}$, and
hence it must be a $c$-number.
Indeed we have
$$
e^{i2\pi L\hat p} = 1.
\eqno(5.7)
$$
This leads to the well-known result,
$$
\hat p = {m\over L} \quad \hbox{\elevenrm with}\ \ m \in
\hbox{\elevenbf Z}.
\eqno(5.8)
$$
In string theory on orbifolds, $\hat p$ and $e^{ik{\hat x}}$
will be replaced by the momentum operators
$P_L^I(z)$,$P_R^I({\bar z})$ and the vertex operator
$V(k_L,k_R;z)$, respectively, and an analog of the identity
(5.7) in the $U$-twisted sector is
$$
\xi_k\  e^{-ik_L\cdot U\hat x_L-ik_R\cdot U\hat x_R}
     \  e^{ik_L\cdot \hat x_L+ik_R\cdot \hat x_R}
     \  e^{i2\pi(k_L\cdot {\cal P}_U\hat p_L-
         k_R\cdot {\cal P}_U\hat p_R)}
      = 1,
\eqno(5.9)
$$
where
$$
\xi_k = \exp \{i\pi (k_L\cdot {\cal P}_Uk_L -
                     k_R\cdot {\cal P}_Uk_R)
               + i{\pi\over 2}(k_L-k_R)\cdot UC_UU^T(k_L-k_R)\}.
\eqno(5.10)
$$
The ${\cal P}_U$ is a projection matrix defined by $(U^N={\bf 1})$
$$
{\cal P}_U = {1\over N}\sum^N_{\ell = 1}U^\ell .
\eqno(5.11)
$$
It is not difficult to show that the left hand side of Eq.(5.9)
commutes with all the operators by use of the commutation
relations (3.9).
A physical meaning of the identity (5.9) is obvious for
$[U,B]=0$.
The identity (5.9) then reduces to
$$
e^{i\pi k\cdot \hat w} e^{i\pi w\cdot U\hat p} = 1,
\eqno(5.12)
$$
where
$$
\eqalign{
w^I &= k^I_L - k^I_R \in \Lambda, \cr
k^I &= (1+B)^{IJ}k^J_L + (1-B)^{IJ}k^J_R \in 2\Lambda^*. \cr
}
\eqno(5.13)
$$
The ${\hat p}^I$ is the canonical momentum conjugate to
${\hat x}^I$ in the $U$-twisted sector
\footnote{$^\dagger$}{\ninerm\baselineskip=11pt
For the precise definition of $\scriptstyle {\hat p}^I$ in the
$\scriptstyle U$
-twisted sector, see Eq.(3.15) in ref.17.}.
The identity (5.12) implies that
$$
\eqalign{
\hat w^I &\in \Lambda, \cr
\hat p^I &\in 2\Lambda^*. \cr
}
\eqno(5.14)
$$
These results are consistent with the identification
$x^I \sim x^I+\pi w^I$ for all $w^I\in\Lambda$.
For $[U,B]\ne 0$, the first relation of (5.14)
still holds while the second one does not.
In fact, putting $k^I=0$ and $w^I\in \Lambda_U$
in Eq.(5.9), we have
$$
\exp \{i\pi w\cdot \hat p - i{\pi\over 2}w\cdot C_Uw\} = 1.
\eqno(5.15)
$$
This is equivalent to Eq. (5.1) or (5.5).

\vglue 0.6cm

\line{\elevenbf 6. Conclusions \hfil}
\vglue 0.4cm

In this paper, we have discussed some of new aspects of string
theory on orbifolds in the presence of the antisymmetric
constant background field $B^{IJ}$ both in a path integral
point of view and in an operator formalism point
of view, although we have omitted some of the details.
This work is, however, far from complete.
We have shown that the transformation (2.4) is a quantum
symmetry of string theory on {\elevenit tori}.
To clarify the full topological nature of the $B^{IJ}$-term,
it is necessary to construct the path integral formalism
of string theory on {\elevenit orbifolds} in the presence of
the antisymmetric background field and also beyond
one-loop.
For topologically nontrivial cases, i.e., $[U,B]\ne 0$, we
have not yet found a Euclidean action of string theory on
orbifolds, which respects the duality and modular invariance
of amplitudes ${}^{17,25}$.
In the operator formalism, we also have to construct Hilbert
spaces of twisted sectors and to verify modular invariance
of amplitudes.
Subtle part for constructing twisted Hilbert spaces is the
zero modes, which obey the nontrivial algebras (3.9).
A necessary condition for modular invariance is left-right
level matching conditions ${}^{26,27}$.
In ref.15, the level matching conditions have been proved to be
satisfied.
The level matching conditions may be necessary and sufficient
for modular invariance as discussed in ref. 26 and 27
although the discussions may not directly be applicable
to our case.
The details of this paper and the subjects addressed above
will be reported elsewhere.

\vglue 0.6cm

\line{\elevenbf Acknowledgements \hfil}
\vglue 0.4cm

I would like to thank K. Ito, H. Ooguri, J. Petersen, F. Ruiz
and M. Tabuse for useful discussions and also would like to
acknowledge the hospitality for the Niels Bohr Institute
where this work was done.

\vglue 0.6cm

\line{\elevenbf References \hfil}
\vglue 0.4cm

\item{1.} I. Frenkel and V. Ka\v c, {\elevenit Invent. Math.}
{\elevenbf 62}
(1980) 23;
\item{  } G. Segal, Commun. {\elevenit Math. Phys.}
{\elevenbf 80} (1981) 301;
\item{2.} P. Goddard and D. Olive, {\elevenit Intern. J. Mod. Phys.}
{\elevenbf A1} (1986) 303.
\item{3.} K. Kikkawa and M. Yamasaki, {\elevenit Phys. Lett.}
{\elevenbf B149} (1984) 357;
\item{  } N. Sakai and I. Senda, {\elevenit Prog. Theor. Phys.}
{\elevenbf 75} (1986) 692.
\item{4.} L. Dixon, J.A. Harvey, C. Vafa and E. Witten,
{\elevenit Nucl. Phys.}
{\elevenbf B261} (1985) 678; {\elevenbf B274} (1986) 285.
\item{5.} I. Frenkel, J. Lepowsky and A. Meurman, Proc. Conf. on
{\elevenit Vertex
operators in mathematics and physics}, ed. J. Lepowsky et al. (Springer
1984).
\item{6.} K.S. Narain, M.H. Sarmadi and C. Vafa, {\elevenit Nucl. Phys.}
{\elevenbf B288} (1987) 551.
\item{7.} K. Kobayashi, {\elevenit Phys. Rev. Lett}.
{\elevenbf 58} (1987) 2507;
\item{  } K. Kobayashi and M. Sakamoto, {\elevenit Z. Phys.}
{\elevenbf C41} (1988) 55;
\item{  } M. Sakamoto, {\elevenit Mod. Phys. Lett.}
{\elevenbf A5} (1990) 1131;
{\elevenit Prog. Theor. Phys.} {\elevenbf 84} (1990) 351.
\item{8.} R. Dijkgraaf, E. Verlinde and H. Verlinde,
{\elevenit Commun. Math. Phys.} {\elevenbf 115} (1988) 649;
\item{    } P. Ginsparg, {\elevenit Nucl. Phys.}
{\elevenbf B295} (1988) 153.
\item{9.} K.S. Narain, M.H. Sarmadi and E. Witten,
{\elevenit Nucl. Phys.} {\elevenbf B279} (1987) 369.
\item{10.} K.S. Narain, {\elevenit Phys. Lett.}
{\elevenbf B169} (1986) 41.
\item{11.} A. Font, L.E. Ib\'a\~ nez, F. Quevedo and A. Sierra,
{\elevenit Nucl. Phys.} {\elevenbf B331} (1990) 421.
\item{12.} J.A. Casas and C. Mu\~ noz,
{\elevenit Nucl. Phys.} {\elevenbf B332} (1990) 189.
\item{13.} Y. Katsuki, Y. Kawamura, T. Kobayashi, N. Ohtsubo, Y. Ono and
K. Tanioka, {\elevenit Nucl. Phys.} {\elevenbf B341} (1990) 611.
\item{14.} A. Fujitsu, T. Kitazoe, M. Tabuse and H. Nishimura,
{\elevenit Intern. J. Mod. Phys.} {\elevenbf A5} (1990) 1529.
\item{15.} M. Sakamoto and M. Tabuse, Kobe preprint KOBE-92-02 (1992);
\item{   } T. Horiguchi, M. Sakamoto and M. Tabuse, Kobe preprint
KOBE-92-03 (1992).
\item{16.} A.M. Polyakov, {\elevenit Phys .Lett.}
{\elevenbf B103} (1981) 207;
\item{   } J. Polchinski, {\elevenit Commun. Math. Phys.}
{\elevenbf 104} (1986) 37.
\item{17.} K. Inoue, S. Nima and H. Takano,
{\elevenit Prog. Theor. Phys.} {\elevenbf 80} (1988) 881.
\item{18.} J. Erler, D. Jungnickel, J. Lauer and J. Mas, preprint
SLAC-PUB-5602 (1991).
\item{19.} Y. Aharonov and D. Bohm, {\elevenit Phys. Rev.}
{\elevenbf 115} (1959) 485.
\item{20.} F. Englert and A. Neveu,
{\elevenit Phys. Lett.} {\elevenbf B163} (1985) 349.
\item{21.} J.H. Schwarz, {\elevenit Phys. Rep.}
{\elevenbf 8C} (1973) 269;
{\elevenbf 89} (1982) 223;
\item{   } J. Scherk, {\elevenit Rev. Mod. Phys.}
{\elevenbf 47} (1975) 123.
\item{22.} V. Ka\v c and D.Peterson, in
{\elevenit Anomalies, geometry and topology,}
ed. A. White (World Scientific, Singapore, 1985);
\item{   } J. Lepowsky, {\elevenit Proc. Nat. Acad. Sci.}
(USA) {\elevenbf 82} (1985) 8295.
\item{23.} M. Sakamoto, {\elevenit Phys. Lett.}
{\elevenbf B231} (1989) 258.
\item{24.} K. Itoh, M. Kato, H. Kunitomo and M. Sakamoto,
{\elevenit Nucl. Phys.}  {\elevenbf B306} (1988) 362.
\item{25.} K. Inoue and S. Nima, {\elevenit Prog. Theor. Phys.}
{\elevenbf 84} (1990) 702.
\item{26.} C. Vafa, {\elevenit Nucl. Phys.}
{\elevenbf B273} (1986) 592.
\item{27.} K.S. Narain, M.H. Sarmadi and C. Vafa,
{\elevenit Nucl. Phys.} {\elevenbf B356} (1991) 163.

\vfill

\eject
\bye